\documentclass[reprint,letterpaper,twocolumn,superscriptaddress,amsmath,amssymb,nofootinbib,preprintnumbers]{revtex4-2} 
 
\usepackage{mathptmx}
\usepackage{graphicx}  
\usepackage[mathlines]{lineno} 
\usepackage[english]{babel} 
\usepackage{xcolor}
\usepackage{ulem}
\usepackage{dcolumn}   
\usepackage{bm}        
\usepackage{amssymb}   
\usepackage{amsbsy}   
\usepackage{hyperref}
\usepackage{nicefrac}
\usepackage{todonotes}
\usepackage{comment}

\hypersetup{
    colorlinks=true,
    linkcolor=cyan,
    filecolor=magenta,      
    urlcolor=cyan,
    citecolor=violet,
}

\makeatletter
\g@addto@macro\bfseries{\boldmath}
\makeatother

\newcommand{\dif}{\mathrm{d}}
\newcommand{\xmax}{X_\text{max}}
\newcommand{\Hinf}{H_\text{inf}}
\newcommand{\Trh}{T_\text{rh}}
\def \LambdaI {\Lambda_{\mathrm{I}}}
\def \nn {\mathbf{n}}

\begin{document}

\title{Limits to gauge coupling in the dark sector set by the non-observation of instanton-\\
induced decay of Super-Heavy Dark Matter in the Pierre Auger Observatory data}
 
\author{P.~Abreu}
\affiliation{Laborat\'orio de Instrumenta\c{c}\~ao e F\'\i{}sica Experimental de Part\'\i{}culas -- LIP and Instituto Superior T\'ecnico -- IST, Universidade de Lisboa -- UL, Lisboa, Portugal}

\author{M.~Aglietta}
\affiliation{Osservatorio Astrofisico di Torino (INAF), Torino, Italy}
\affiliation{INFN, Sezione di Torino, Torino, Italy}

\author{J.M.~Albury}
\affiliation{University of Adelaide, Adelaide, S.A., Australia}

\author{I.~Allekotte}
\affiliation{Centro At\'omico Bariloche and Instituto Balseiro (CNEA-UNCuyo-CONICET), San Carlos de Bariloche, Argentina}

\author{K.~Almeida Cheminant}
\affiliation{Institute of Nuclear Physics PAN, Krakow, Poland}

\author{A.~Almela}
\affiliation{Instituto de Tecnolog\'\i{}as en Detecci\'on y Astropart\'\i{}culas (CNEA, CONICET, UNSAM), Buenos Aires, Argentina}
\affiliation{Universidad Tecnol\'ogica Nacional -- Facultad Regional Buenos Aires, Buenos Aires, Argentina}

\author{R.~Aloisio}
\affiliation{Gran Sasso Science Institute, L'Aquila, Italy}
\affiliation{INFN Laboratori Nazionali del Gran Sasso, Assergi (L'Aquila), Italy}

\author{J.~Alvarez-Mu\~niz}
\affiliation{Instituto Galego de F\'\i{}sica de Altas Enerx\'\i{}as (IGFAE), Universidade de Santiago de Compostela, Santiago de Compostela, Spain}

\author{R.~Alves Batista}
\affiliation{IMAPP, Radboud University Nijmegen, Nijmegen, The Netherlands}

\author{J.~Ammerman Yebra}
\affiliation{Instituto Galego de F\'\i{}sica de Altas Enerx\'\i{}as (IGFAE), Universidade de Santiago de Compostela, Santiago de Compostela, Spain}

% ...

\collaboration{The Pierre Auger Collaboration}
\email{spokespersons@auger.org}
\homepage{http://www.auger.org}
\noaffiliation



\begin{abstract}

\noindent Instantons, which are non-perturbative solutions to Yang-Mills equations, provide a signal for the occurrence of quantum tunneling between distinct classes of vacua. They can give rise to decays of particles otherwise forbidden. Using data collected at the Pierre Auger Observatory, we search for signatures of such instanton-induced processes that would be suggestive of super-heavy particles decaying in the Galactic halo. These particles could have been produced during the post-inflationary epoch and match the relic abundance of dark matter inferred today. The non-observation of the signatures searched for allows us to derive a bound on the reduced coupling constant of gauge interactions in the dark sector: $\alpha_X \alt 0.09$, for $10^{9} \alt M_X/{\rm GeV} < 10^{19}$. Conversely, we obtain that, for instance, a reduced coupling constant $\alpha_X = 0.09$ excludes masses $M_X \gtrsim 3\times 10^{13}~$GeV. In the context of dark matter production from gravitational interactions alone, we illustrate how these bounds are complementary to those obtained on the Hubble rate at the end of inflation from the non-observation of tensor modes in the cosmological microwave background.

\end{abstract}

\maketitle

\begin{figure}
\centering
\includegraphics[width=\columnwidth]{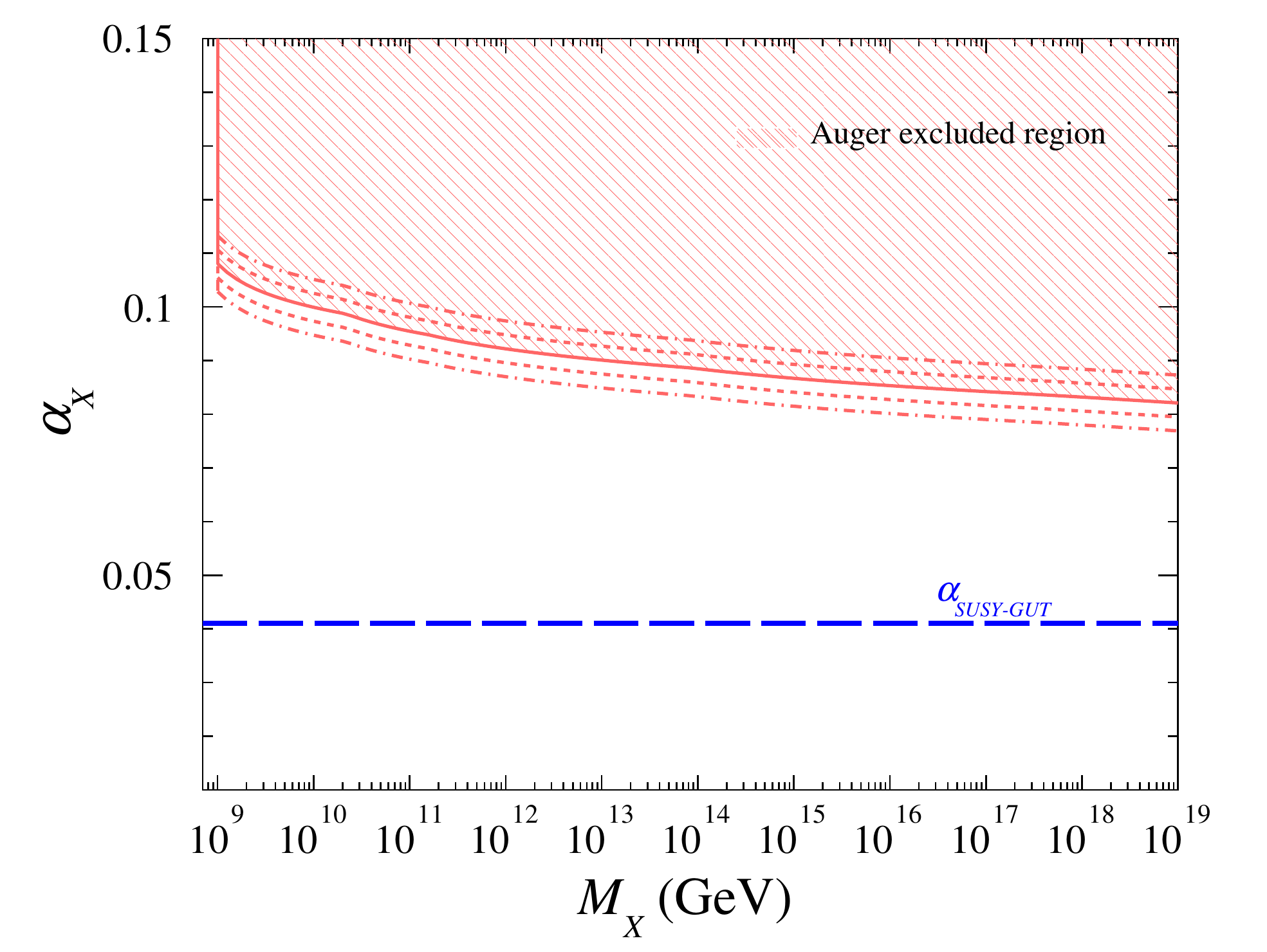}
\caption{Upper limits at 95\% C.L.\ on the  coupling constant $\alpha_X$ of a hidden gauge interaction as a function of the mass $M_X$ of a dark matter particle decaying into a dozen $q \bar q$ pairs. For reference, the unification of the three SM gauge couplings is shown as the blue dashed line in the framework of supersymmetric GUT~\cite{10.1093/ptep/ptaa104}.}
\label{fig:alphaX-mass}
\end{figure}

Should a flux of astrophysical photons with energies in excess of ${\simeq}~10^8$\,GeV be detected, it could be compelling evidence for the decay of super-heavy relics dating from the early universe~\cite{Bhattacharjee:1999mup,Anchordoqui:2018qom}. Possible mechanisms taking place during or at the end of the inflationary era in Big Bang cosmology have been shown to be capable of producing such particles~\cite{Ellis:1990nb,PhysRevLett.79.4302,Chung:1998zb,Kuzmin:1998kk,Chung:1999ve,Fedderke:2014ura,Garny:2015sjg,Ellis:2015jpg,Kolb:2017jvz,Dudas:2017rpa,Kaneta:2019zgw,Mambrini:2021zpp}. The abundance of the stable super-heavy particles could then evolve to match the relic abundance of dark matter (DM) inferred today, for viable parameters governing the thermal history and the geometry of the universe, such as the reheating temperature or the Hubble expansion rate at the end of inflation. Stability for super-heavy particles is more easily achieved for a dark sector interacting with the standard model (SM) sector only via gravity. The absence of other DM-SM couplings is consistent with the extensive observational evidence for the existence of DM based on gravitational effects alone. However, even particles protected from decay by a symmetry can eventually disintegrate due to non-perturbative effects in non-abelian gauge theories and produce ultra-high energy (UHE) photons. In this Letter, we show that the absence of such photons in the data of the Pierre Auger Observatory provides constraints on the coupling constant of a hidden sector pertaining to super-heavy dark matter (SHDM), possibly unified with SM interactions at a high scale. The constraints are illustrated in Fig.~\ref{fig:alphaX-mass} in terms of the reduced coupling constant of a hidden gauge interaction and the mass of the SHDM candidate. Our results show that the coupling should be less than $\simeq 0.09$ for a wide range of masses. After explaining how these constraints are obtained, we briefly discuss their relevance for delineating viable regions of cosmological parameters, in a manner complementary to the constraints provided by the non-detection so far of tensor modes in the cosmological microwave background anisotropies~\cite{BICEP2:2015nss,Planck:2015sxf}.
\\

\textit{Contemporary motivations for SHDM.} Technical naturalness has provided an important motive for the emergence of new physics at the TeV scale~\cite{tHooft:1979rat}, but the corresponding new particles have escaped detection so far~\cite{MarrodanUndagoitia:2015veg,Rappoccio:2018qxp,Gaskins:2016cha}. An alternative tool to infer the energy scale of new physics relies on assessing the scale $\LambdaI$ at which the Higgs potential develops an instability at large field values. Its estimation at the two-loop level was made possible by the precise measurements of the Higgs mass and the top Yukawa coupling~\cite{Buttazzo:2013uya,Alekhin:2012py,Bednyakov:2015sca}. It turns out to result in a very high energy scale, $\LambdaI=10^{10}$ to $10^{12}$\,GeV. Moreover, the particular slow running of the Callan-Symanzik $\beta_\lambda$ function relative to the self-Higgs coupling makes it possible to extrapolate the SM up to $M_\text{Pl}$ without encountering any instabilities~\cite{Buttazzo:2013uya}. Renouncing naturalness to solve the problem of the mass hierarchy, new degrees of freedom could thus appear only in the range between $\LambdaI$ and $M_\text{Pl}$, motivating searches for SHDM. We note also that while some have argued that the properties of nuclei and atoms would not allow complex chemistry if the electroweak scale were too far from the confinement scale of QCD~\cite{Damour:2007uv}, there is no such anthropic requirement for the mass scale of DM. 

Independent of the intrinsic consistency of the SM up to a very-high scale, models of gravitational production of DM provide another motivation for a spectrum of super-heavy particles. While no coupling between the SM and DM sectors is introduced in the concordance model of cosmology, most DM models invoke some weak couplings, or new feeble couplings, to explain DM production during the post-inflationary reheating period. It turns out, however, that the introduction of such couplings is not a compelling necessity if one considers the minimal assumption of graviton exchange to act as the only portal. Recent studies have indeed shown that, on the condition that DM is super heavy, the relic abundance observed today can be reproduced for tenable ranges of quantities governing the inflationary and reheating eras in the early Universe~\cite{Garny:2015sjg,Mambrini:2021zpp}. In addition, while structure formation constrains the mass density of DM, it does not preclude SHDM models as it leaves a \textit{carte blanche} for the mass spectrum of the particles.

SHDM particles interacting with SM particles through gravitons alone have been dubbed as Planckian-interacting massive particles (PIDM)~\cite{Garny:2015sjg}, and we shall use this term when we need to be specific. There are only a few possible signatures to test this scenario for DM. We show that if instanton effects are strong enough, PIDM particles could decay and their by-products could be detected in ultra-high energy cosmic ray (UHECR) data. Conversely, the non-observation of these by-products allows us to set upper bounds on the dark sector coupling constant. We note that these limits are, to date, the best ever obtained from instanton-mediated processes; they are an indirect probe of the instanton strength.  \\

\textit{Decay mechanisms of SHDM particles.} Some SHDM models postulate the existence of super-weak couplings between the dark and SM sectors. The lifetime $\tau_X$ of the particles is then governed by the strength of the couplings $g_X$ and by the dimension $n$ of the operator standing for the SM fields in the effective interaction~\cite{deVega:2003hh}. This results in lifetimes that are in general far too short for DM to be stable enough, unless a practically untenable fine tuning between $g_X$ and $n$ holds~\cite{Ellis:1990nb,deVega:2003hh,PierreAuger:2022ibr}. Stability of super-heavy particles is thus preferentially calling for a new quantum number conserved in the dark sector so as to protect the particles from decaying. Nevertheless, as we have already pointed out in the study motivation, even stable particles in the perturbative domain will in general eventually decay due to non-perturbative effects in non-abelian gauge theories~\cite{Belavin:1975fg,Coleman:1978ae,Vainshtein:1981wh}. 

Instanton-induced decay can thus make observable a dark sector that would otherwise be totally hidden by the conservation of a quantum number~\cite{Kuzmin:1997jua}. Assuming quarks and leptons carry this quantum number and so contribute to anomaly relationships with contributions from the dark sector, they will be by-product of decays together with the lightest hidden fermion. The lifetime of the decaying particle follows from Ref.~\cite{tHooft:1976rip},
\begin{equation}
\label{eqn:tauX}
\tau_X \simeq M_X^{-1}\exp{\left(4\pi/\alpha_X\right)},
\end{equation}
with $\alpha_X$ being the reduced coupling constant of the hidden gauge interaction. In this expression, we retained only the exponential dependency in $\alpha_X^{-1}$, dropping the functional determinants 
arising from the exact content of fields of the underlying theory. The constraints inferred using Eq.~\eqref{eqn:tauX} are indeed barely destabilized for a wide range of numerical factors given the exponential dependency in $\alpha_X^{-1}$. Eq.~\eqref{eqn:tauX} provides us with a relationship connecting the lifetime $\tau_X$, which is shown below to be constrained by the absence of UHE photons, to the coupling constant $\alpha_X$. 
\\

\textit{Production of ultra-high energy photons.} In most SHDM models, the production of quark/anti-quark pairs is expected in the decay by-products, giving rise to large fluxes of UHE particles such as nucleons, photons and neutrinos. This is because each pair triggers a QCD cascade until the hadronization of the partons occurs and the unstable hadrons eventually decay. Various computational schemes have been applied to predict the energy spectra of the UHE particles~\cite{Aloisio:2003xj,Sarkar:2001se,Barbot:2002gt,Kachelriess:2018rty,Alcantara:2019sco}. The fragmentation of a parton into a hadron is determined from the fragmentation functions of partons convolved with hadronization functions, which do not depend on the scale $M_X$ and can therefore be calculated from the available data. The fragmentation functions, on the other hand, are evolved starting from measurements at the electroweak scale up to the energy scale fixed by $M_X$ using the DGLAP equation. We use the scheme detailed in Ref.~\cite{Aloisio:2003xj} in this study. Overall, the spectra of the UHE particles is of the form $E^{-1.9}$ in the $q \bar q$ channel, and is barely softened by kinematical effects in large-multiplicity final states~\cite{Sarkar:2001se,Barbot:2002et}.
 
As shown below, it turns out to be more efficient to search for decaying super-heavy particles via UHE-photon by-products. Due to their attenuation over intergalactic distances, only those emitted in the Milky Way can survive on their way to Earth. The emission rate per unit volume and unit energy $q_\gamma$ from any point labelled by its Galactic coordinates is shaped by the density of DM particles, $n_\text{DM}$,
\begin{equation}
    \label{eqn:q_gamma}
    q_\gamma(E,\mathbf{x}_\odot+s\nn) = \frac{1}{\tau_X}\frac{\dif N_\gamma}{\dif E}n_\text{DM}(\mathbf{x}_\odot+s\nn),
\end{equation}
where $\tau_X$ is the lifetime of the particle, $\mathbf{x}_\odot$ is the position of the Solar system in the Galaxy, $s$ is the distance from $\mathbf{x}_\odot$ to the emission point, and $\nn\equiv\nn(\ell,b)$ is a unit vector on the sphere pointing to the Galactic longitude $\ell$ and latitude $b$. Hereafter, the density is more conveniently expressed in terms of energy density $\rho_\text{DM}(\mathbf{x})=M_Xn_\text{DM}(\mathbf{x})$, normalized to $\rho(\mathbf{x}_\odot)=0.3$\,GeV\,cm$^{-3}$. There are uncertainties in the determination of this profile. We take as reference the traditional NFW profile~\cite{Navarro:1995iw}. We have checked that other profiles such as those from Einasto~\cite{Einasto:1965czb}, Burkert~\cite{Burkert:1995yz} or Moore~\cite{Moore:1999nt} would lead to differences within 2\% in the determination of $\alpha_X$. 

The expected flux (per steradian) of UHE photons produced by the decay of super-heavy particles, $J_{\text{DM},\gamma}(E,\nn)$, is obtained by integrating the position-dependent emission rate $q_\gamma$ along the path of the photons in the direction $\nn$,
\begin{equation}
    \label{eqn:Jgal}
    J_{\mathrm{DM},\gamma}(E,\nn)=\frac{1}{4\pi}\int_0^\infty \dif s~q_\gamma(E,\mathbf{x}_{\odot}+s\nn),
\end{equation}
where the $4\pi$ normalization factor accounts for the isotropy of the decay processes. While the peak value of the flux is inversely proportional to the unknown $M_X$ and $\tau_X$ parameters, the energy and directional dependencies are determined by Eq.~\eqref{eqn:q_gamma}.
The exact content of quarks and leptons  depends on the specific underlying model. Yet, instanton-induced decays obey selection rules that involve necessarily large multiplicities. As a proxy, we consider a dozen $q\overline{q}$ pairs and adapt $\dif N_\gamma/\dif E$ in Eq.~\eqref{eqn:q_gamma} accordingly~\cite{PierreAuger:2022ibr}. The flux pattern on the sky is more intense in a hot-spot region around the Galactic center; it provides therefore clear signatures. On the other hand, the non-observation of UHE photons enables one to constrain the all-sky flux observed over the solid angle $\Delta\Omega$, $\langle J_{\mathrm{DM},\gamma}(E,\nn)\rangle=\int_{\Delta\Omega}\dif\nn ~J_{\mathrm{DM},\gamma}(E,\nn)/\Delta\Omega$, and thus to constrain the unknown $M_X$ and $\tau_X$ parameters. \\

\textit{Constraints on dark-sector coupling constant from instanton-induced decays.} Of particular interest would thus be the detection of UHE photons from regions of denser DM density such as the center of our Galaxy. Due to the spectral steepness of the expected flux, this search can presently only be done through large ground-based detectors that exploit the phenomenon of extensive air showers. The identification of photon primaries relies on the ability to distinguish showers generated by photons from those initiated by the overwhelming background of protons and heavier nuclei. Since the radiation length in the atmosphere is more than two orders of magnitude smaller than the mean free path for photo-nuclear interactions, the transfer of energy to the hadron/muon channel is reduced in photon showers with respect to the bulk of hadron-induced showers, resulting in a lower number of secondary muons. Additionally, as the development of photon showers is delayed by the typically small multiplicity of electromagnetic interactions, they reach the maximum development of the shower, $\xmax$, deeper in the atmosphere with respect to showers initiated by hadrons. 

Both the ground signal and $\xmax$ can be measured at the Pierre Auger Observatory~\cite{PierreAuger:2015eyc}, where a hybrid detection technique is employed for the observation of extensive air showers by combining a fluorescence detector (FD) with a ground array of particle detectors (surface detector, SD) separated by 1500\,m. The FD provides direct observation of the longitudinal shower profile, which allows for the measurement of the energy and the $\xmax$ of a shower, while the SD samples the secondary particles at ground level. Although showers are observed at a fixed slice in depth with the SD, the longitudinal development is embedded in the signals detected. The FD and SD are complemented with the low-energy enhancements of the Observatory, namely three additional fluorescence telescopes with an elevated field of view, overlooking a denser SD array, in which the stations are separated by 750~m. The combination of these instruments allows showers to be measured in the energy range above $10^{8}$\,GeV. 

\begin{figure}
\centering
\includegraphics[width=\columnwidth]{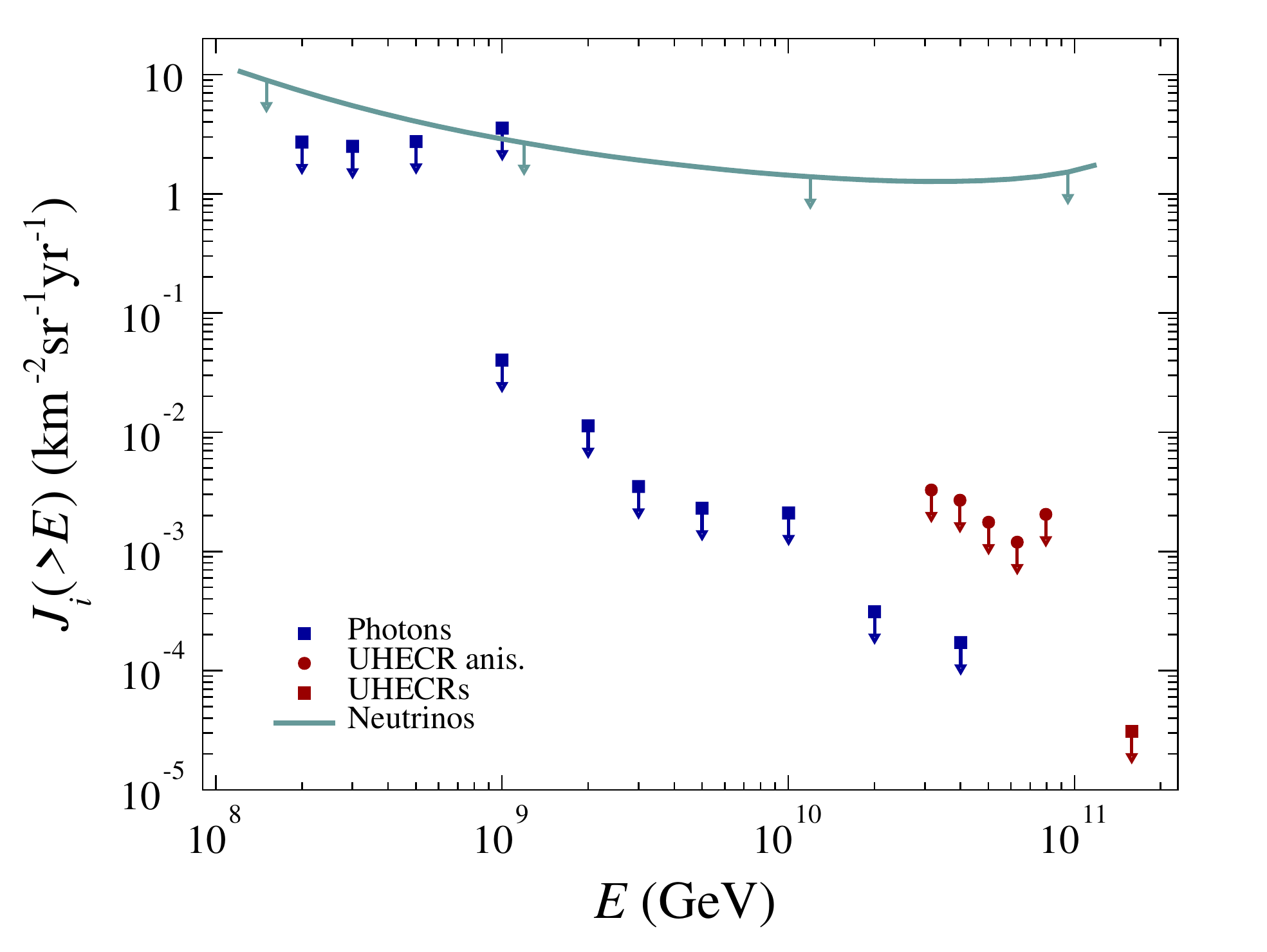}
\caption{Flux upper limits of UHE photons, neutrinos and cosmic rays as a function of energy thresholds.}
\label{fig:upplim}
\end{figure}

Three different analyses, differing in the detector used, have been developed to cover the wide energy range probed at the Observatory~\cite{PierreAuger:2022uwd,Savina:2021ufs,PierreAuger:2022aty}. No photons with energies above $2{\times}10^{8}$\,GeV have been unambiguously identified so far, leading to the 95\% C.L.\ flux upper limits displayed in Fig.~\ref{fig:upplim}. The limit above $10^{11.2}$\,GeV, stemming from the non-detection so far of any UHECR~\cite{PierreAuger:2020qqz}, including photons, is also constraining~\cite{Alcantara:2019sco,Anchordoqui:2021crl}. For comparison purposes, neutrino limits obtained at the Observatory~\cite{PierreAuger:2019ens} are also displayed as the continuous line. Indeed, neutrinos constitute in the figure another emblematic signature of decays of super-heavy particles. Except at the lowest energies, these limits are seen to be superseded by photon limits, as are those from anisotropy signatures searched for in the bulk of UHECR data shown as red-filled circles~\cite{PierreAuger:2022ibr}. 

Assuming that the relic abundance of DM is saturated by super-heavy particles, constraints can be inferred in the plane $(\tau_X,M_X)$ by requiring the all-sky flux integrated above some energy threshold $E$ to be less than the limits, $\int_E^\infty\dif E'\langle J_{\text{DM},\gamma}(E',\nn)\rangle \leq J^{95\%}_\gamma({\geq}E)$. For a specific upper limit at one energy threshold, a scan of the value of the mass $M_X$ is carried out so as to infer a lower limit of the $\tau_X$ parameter, which is subsequently transformed into an upper limit on $\alpha_X$ by means of Eq.~\eqref{eqn:tauX}. This defines a curve. By repeating the procedure for each upper limit on $J^{95\%}_\gamma({\geq}E)$, a set of curves is obtained, reflecting the sensitivity of a specific energy threshold to some range of mass. The union of the excluded regions finally provides the constraints in the plane $(\alpha_X,M_X)$. In this manner the shaded red area is obtained in Fig.~\ref{fig:alphaX-mass}. As already noted, additional model-dependent factors could be at play in the vacuum transition amplitude~\cite{tHooft:1976snw} and thus in Eq.~\eqref{eqn:tauX}. Explicit constructions of the dark sector are required to calculate these factors. Such constructions are well beyond the scope of this study. Although the limits presented in Fig.~\ref{fig:alphaX-mass} are hardly destabilized due to the exponential dependence in $\alpha_X^{-1}$, we note that a shift of $\pm 0.0013k$ would arise for factors $10^{\pm k}$. We limit ourselves to showing with dotted and dashed lines the bounds that would be obtained for $k=2$ and $k=4$, respectively. These factors are by far the dominant systematic uncertainties.
\\ 
 
\textit{Connection to cosmological scenarios.} We now briefly mention how the results shown in Fig.~\ref{fig:alphaX-mass} can be connected to cosmological scenarios. Further details can be found in a companion paper~\cite{PierreAuger:2022ibr}. In inflationary cosmologies, the inflaton field responsible for the rapid expansion of the Universe, $\phi$, slowly rolls down to its minimum of potential before starting to oscillate about this minimum. This marks the end of the inflation era at time $\Hinf^{-1}$, with $\Hinf$ the Hubble rate at this time, and the beginning of a matter-dominated era during which the production of SM particles accompanying the decay of coherent oscillations of the inflaton field reheats the Universe. The temperature rises rapidly to a maximum before decreasing slowly until the reheating era ends at time $\Gamma_\phi^{-1}$, marking the beginning of the radiation-dominated era when the temperature decreases more rapidly as $a^{-1}$, with $a$ being the cosmological scale factor. The temperature at the end of reheating, $\Trh$, is, together with $\Hinf$, an important parameter governing the dynamics of the reheating era summarized here -- see Ref.~\cite{Giudice:2000ex} for details. A relevant combination of these parameters is the reheating efficiency $\epsilon$, which, defined as $\epsilon=(\Gamma_\phi/\Hinf)^{1/2}$~\cite{Kolb:1990vq}, measures the duration of the reheating period. It can be related to $\Trh$ and $\Hinf$ through $\epsilon\simeq  \Trh/(0.25(M_\text{Pl}\Hinf)^{1/2})$~\cite{Garny:2015sjg}.

PIDM particles can be produced during reheating by annihilation of SM particles~\cite{Garny:2015sjg} or inflaton particles~\cite{Mambrini:2021zpp} through gravitational interaction. The energy density of the universe is then in the form of unstable inflaton particles, SM radiation and stable massive particles, the time evolution of which is governed by a set of coupled Boltzmann equations~\cite{Chung:1998rq}. However, because the energy density of the massive particles is always sub-dominant, the evolution of the  inflationary and radiation energy densities largely decouple from the time evolution of the PIDM-particle density $n_X$. In addition, because PIDM particles interact through gravitation only, they never come to thermal equilibrium. In this case, the collision term in the Boltzmann equation can be approximated as a source term only,
\begin{equation}
\label{eqn:nX}
\frac{\dif n_X(t)}{\dif t}+3H(t)n_X(t)\simeq \sum_i \gamma_i\overline{n}_i^2(t).
\end{equation}
Here, the sum on the right hand side represents the contributions from the SM and inflationary sectors. Using, on the one hand, the evolution of the SM-matter and inflaton densities derived in Ref.~\cite{Chung:1998zb} and Ref.~\cite{Mambrini:2021zpp} respectively, and, on the other hand, the SM+SM$\to$PIDM+PIDM and $\phi+\phi\to$PIDM+PIDM reaction rates derived for fermionic DM in Ref.~\cite{Garny:2017kha} and Ref.~\cite{Mambrini:2021zpp} respectively, the present-day relic abundance of DM, $\Omega_X$, can be related to $M_X$, $H_\text{inf}$, and $\epsilon$ through
\begin{equation}
\label{eqn:abundance}
    \Omega_Xh^2=\frac{1.4{\times}10^{23}\epsilon M_X}{M^{5/2}_\text{Pl}\Hinf^{3/2}}\int_{a_\text{inf}}^\infty\frac{\dif a}{H(a)}\sum_i \gamma_ia^2\overline{n}_i^2(a),
\end{equation}
where $h$ is the dimensionless expansion rate and $a_\text{inf}$ the scale factor at the end of inflation.

\begin{figure}
\centering
\includegraphics[width=\columnwidth]{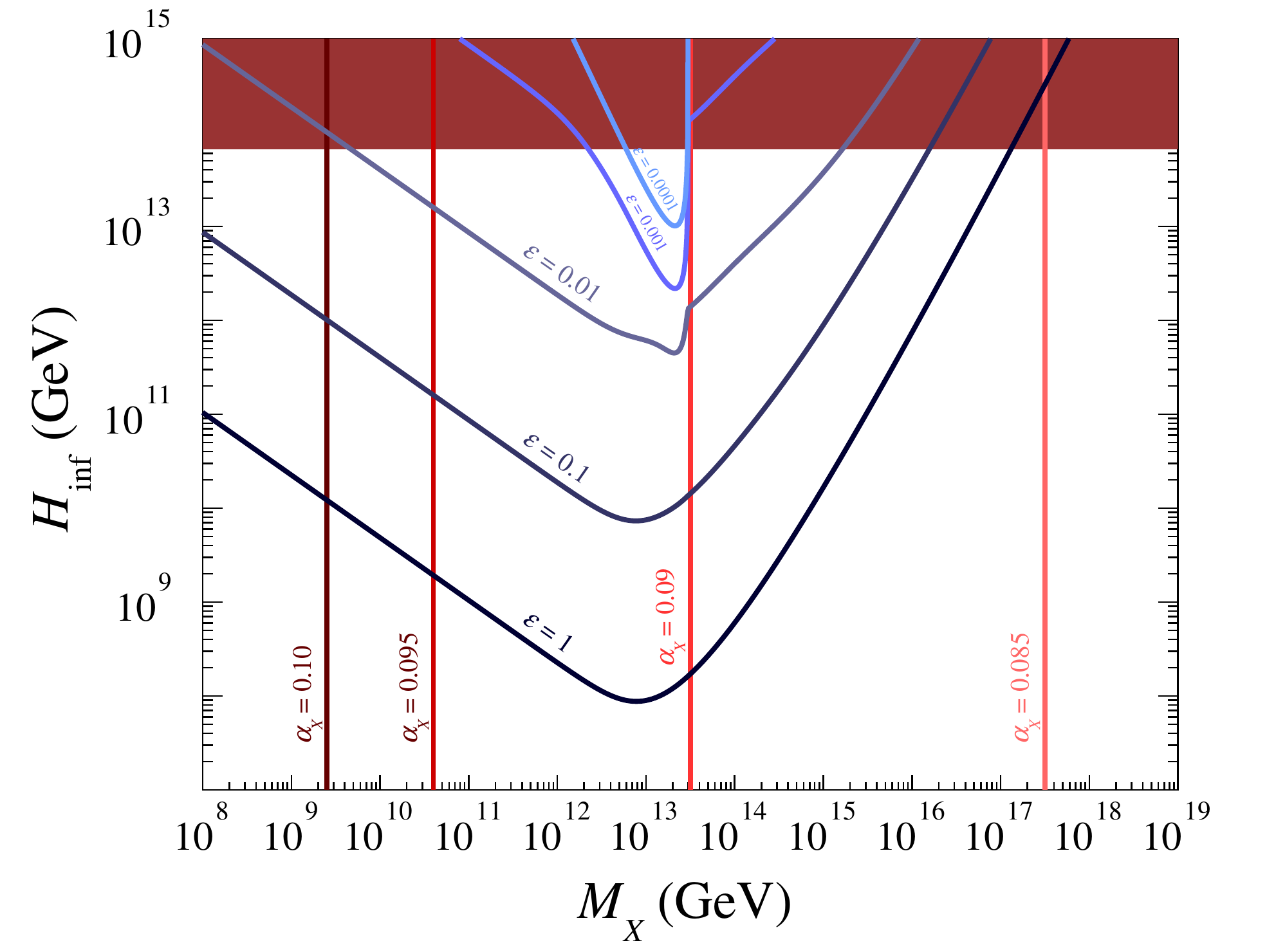}
\caption{Constraints in the $(H_\text{inf},M_X)$ plane. The red  region is excluded by the non-observation of tensor modes in the cosmic microwave background~\cite{Garny:2015sjg,Planck:2015sxf}. The regions of viable $(H_\text{inf},M_X)$ values needed to set the right abundance of DM are delineated by the blue lines for different values of reheating efficiency $\epsilon$~\cite{Garny:2017kha}. Additional constraints from the non-observation of instanton-induced decay of SHDM particles allow for excluding the mass ranges in the regions to the right of the vertical lines, for the specified values of the dark-sector gauge coupling.}
\label{fig:HMX}
\end{figure}

The viable $(H_\text{inf},M_X)$ parameter space satisfying Eq.~\eqref{eqn:abundance} is delineated by the blue curves corresponding to different values of $\epsilon$ in Fig.~\ref{fig:HMX}. Values for $(H_\text{inf},M_X)$ above (below) the lines lead to overabundance of (negligible quantity of) DM. Arbitrarily large values of $H_\text{inf}$ are however not permitted because of the 95\% C.L.\ on the tensor-to-scalar ratio in the cosmic microwave background anisotropies, which, once converted into limits on the energy scale of inflation when the pivot scale exits the Hubble radius, yield $H_\text{inf}\leq 4.9{\times}10^{-6}M_\text{Pl}$~\cite{Garny:2015sjg,Planck:2015sxf}. For efficiencies larger than a few percent, PIDM particles are dominantly produced by the thermal bath of SM particles. A wide range of masses $M_X$ is then allowed, including the Grand Unified scale, provided that the energy scale of the inflation ($H_\text{inf}$ being the proxy) is high enough~\cite{Garny:2015sjg} and that the dark-sector gauge coupling $\alpha_X$ is less than $\simeq 0.085$. Larger values of $\alpha_X$ shrink the allowed range of $M_X$, with, for instance, $M_X\alt 2\times 10^9~$GeV for $\alpha_X=0.1$. For efficiencies below the percent level, the production of PIDM particles from the inflaton condensate dominates, allowing smaller values of $\Trh$ to be viable. The allowed region for $M_X$ shrinks around $10^{13}~$GeV, close to the inflaton mass adopted here ($3\times 10^{13}~$GeV). We see that the scenario is then tenable for $\alpha_X \alt 0.09$.

In summary, we have illustrated here the power of upper limits on the flux of UHE photons obtained at the Pierre Auger Observatory to place constraints on physics in the reheating epoch that could be related to Grand Unified models. The minimal setup to produce DM is from gravitational effects alone, consistent with the concordance model of cosmology. This production mechanism could lead to high values of the Hubble rate at the end of inflation that could be revealed by future measurements of primordial tensor-to-scalar ratio provided that $\Hinf\gtrsim 6\times 10^{12}~$GeV~\cite{CMB-S4:2016ple,Ishino:2016wxl}. However, the only unambiguous signature to capture the existence of PIDM is through the detection of UHE photons produced by the instanton-induced decay. The non-observation of such fluxes has allowed us to probe in a unique way to date the instanton strength through the dark-sector gauge coupling. It is likely that the use of limits on UHE photon fluxes made in this Letter only scratches the surface of the power of these limits to constrain physics otherwise beyond the reach of laboratory experiments. 
\\

\textit{Acknowledgments.} The successful installation, commissioning, and operation of the Pierre Auger Observatory would not have been possible without the strong commitment and effort from the technical and administrative staff in Malarg\"ue. We are very grateful to the following agencies and organizations for financial support:
Argentina -- Comisi\'on Nacional de Energ\'\i{}a At\'omica; Agencia Nacional de
Promoci\'on Cient\'\i{}fica y Tecnol\'ogica (ANPCyT); Consejo Nacional de
Investigaciones Cient\'\i{}ficas y T\'ecnicas (CONICET); Gobierno de la
Provincia de Mendoza; Municipalidad de Malarg\"ue; NDM Holdings and Valle
Las Le\~nas; in gratitude for their continuing cooperation over land
access; Australia -- the Australian Research Council; Belgium -- Fonds
de la Recherche Scientifique (FNRS); Research Foundation Flanders (FWO);
Brazil -- Conselho Nacional de Desenvolvimento Cient\'\i{}fico e Tecnol\'ogico
(CNPq); Financiadora de Estudos e Projetos (FINEP); Funda\c{c}\~ao de Amparo \`a
Pesquisa do Estado de Rio de Janeiro (FAPERJ); S\~ao Paulo Research
Foundation (FAPESP) Grants No.~2019/10151-2, No.~2010/07359-6 and
No.~1999/05404-3; Minist\'erio da Ci\^encia, Tecnologia, Inova\c{c}\~oes e
Comunica\c{c}\~oes (MCTIC); Czech Republic -- Grant No.~MSMT CR LTT18004,
LM2015038, LM2018102, CZ.02.1.01/0.0/0.0/16{\textunderscore}013/0001402,
CZ.02.1.01/0.0/0.0/18{\textunderscore}046/0016010 and
CZ.02.1.01/0.0/0.0/17{\textunderscore}049/0008422; France -- Centre de Calcul
IN2P3/CNRS; Centre National de la Recherche Scientifique (CNRS); Conseil
R\'egional Ile-de-France; D\'epartement Physique Nucl\'eaire et Corpusculaire
(PNC-IN2P3/CNRS); D\'epartement Sciences de l'Univers (SDU-INSU/CNRS);
Institut Lagrange de Paris (ILP) Grant No.~LABEX ANR-10-LABX-63 within
the Investissements d'Avenir Programme Grant No.~ANR-11-IDEX-0004-02;
Germany -- Bundesministerium f\"ur Bildung und Forschung (BMBF); Deutsche
Forschungsgemeinschaft (DFG); Finanzministerium Baden-W\"urttemberg;
Helmholtz Alliance for Astroparticle Physics (HAP);
Helmholtz-Gemeinschaft Deutscher Forschungszentren (HGF); Ministerium
f\"ur Innovation, Wissenschaft und Forschung des Landes
Nordrhein-Westfalen; Ministerium f\"ur Wissenschaft, Forschung und Kunst
des Landes Baden-W\"urttemberg; Italy -- Istituto Nazionale di Fisica
Nucleare (INFN); Istituto Nazionale di Astrofisica (INAF); Ministero
dell'Istruzione, dell'Universit\'a e della Ricerca (MIUR); CETEMPS Center
of Excellence; Ministero degli Affari Esteri (MAE); M\'exico -- Consejo
Nacional de Ciencia y Tecnolog\'\i{}a (CONACYT) No.~167733; Universidad
Nacional Aut\'onoma de M\'exico (UNAM); PAPIIT DGAPA-UNAM; The Netherlands
-- Ministry of Education, Culture and Science; Netherlands Organisation
for Scientific Research (NWO); Dutch national e-infrastructure with the
support of SURF Cooperative; Poland -- Ministry of Education and
Science, grant No.~DIR/WK/2018/11; National Science Centre, Grants
No.~2016/22/M/ST9/00198, 2016/23/B/ST9/01635, and 2020/39/B/ST9/01398;
Portugal -- Portuguese national funds and FEDER funds within Programa
Operacional Factores de Competitividade through Funda\c{c}\~ao para a Ci\^encia
e a Tecnologia (COMPETE); Romania -- Ministry of Research, Innovation
and Digitization, CNCS/CCCDI -- UEFISCDI, projects PN19150201/16N/2019,
PN1906010, TE128 and PED289, within PNCDI III; Slovenia -- Slovenian
Research Agency, grants P1-0031, P1-0385, I0-0033, N1-0111; Spain --
Ministerio de Econom\'\i{}a, Industria y Competitividad (FPA2017-85114-P and
PID2019-104676GB-C32), Xunta de Galicia (ED431C 2017/07), Junta de
Andaluc\'\i{}a (SOMM17/6104/UGR, P18-FR-4314) Feder Funds, RENATA Red
Nacional Tem\'atica de Astropart\'\i{}culas (FPA2015-68783-REDT) and Mar\'\i{}a de
Maeztu Unit of Excellence (MDM-2016-0692); USA -- Department of Energy,
Contracts No.~DE-AC02-07CH11359, No.~DE-FR02-04ER41300,
No.~DE-FG02-99ER41107 and No.~DE-SC0011689; National Science Foundation,
Grant No.~0450696; The Grainger Foundation; Marie Curie-IRSES/EPLANET;
European Particle Physics Latin American Network; and UNESCO.

We acknowledge for this work the support of the Institut Pascal at Universit\'e Paris-Saclay during the Paris-Saclay Astroparticle Symposium 2021, with the support of the P2IO Laboratory of Excellence (program ``Investissements d’avenir'' ANR-11-IDEX-0003-01 Paris-Saclay and ANR-10-LABX-0038), the P2I axis of the Graduate School Physics of Universit\'e Paris-Saclay, as well as IJCLab, CEA, IPhT, APPEC, the IN2P3 master projet UCMN and EuCAPT ANR-11-IDEX-0003-01 Paris-Saclay and ANR-10-LABX-0038).

\bibliographystyle{apsrev4-1}
\bibliography{biblio}

\begin{thebibliography}{56}%
\makeatletter
\providecommand \@ifxundefined [1]{%
 \@ifx{#1\undefined}
}%
\providecommand \@ifnum [1]{%
 \ifnum #1\expandafter \@firstoftwo
 \else \expandafter \@secondoftwo
 \fi
}%
\providecommand \@ifx [1]{%
 \ifx #1\expandafter \@firstoftwo
 \else \expandafter \@secondoftwo
 \fi
}%
\providecommand \natexlab [1]{#1}%
\providecommand \enquote  [1]{``#1''}%
\providecommand \bibnamefont  [1]{#1}%
\providecommand \bibfnamefont [1]{#1}%
\providecommand \citenamefont [1]{#1}%
\providecommand \href@noop [0]{\@secondoftwo}%
\providecommand \href [0]{\begingroup \@sanitize@url \@href}%
\providecommand \@href[1]{\@@startlink{#1}\@@href}%
\providecommand \@@href[1]{\endgroup#1\@@endlink}%
\providecommand \@sanitize@url [0]{\catcode `\\12\catcode `\$12\catcode
  `\&12\catcode `\#12\catcode `\^12\catcode `\_12\catcode `\%12\relax}%
\providecommand \@@startlink[1]{}%
\providecommand \@@endlink[0]{}%
\providecommand \url  [0]{\begingroup\@sanitize@url \@url }%
\providecommand \@url [1]{\endgroup\@href {#1}{\urlprefix }}%
\providecommand \urlprefix  [0]{URL }%
\providecommand \Eprint [0]{\href }%
\providecommand \doibase [0]{http://dx.doi.org/}%
\providecommand \selectlanguage [0]{\@gobble}%
\providecommand \bibinfo  [0]{\@secondoftwo}%
\providecommand \bibfield  [0]{\@secondoftwo}%
\providecommand \translation [1]{[#1]}%
\providecommand \BibitemOpen [0]{}%
\providecommand \bibitemStop [0]{}%
\providecommand \bibitemNoStop [0]{.\EOS\space}%
\providecommand \EOS [0]{\spacefactor3000\relax}%
\providecommand \BibitemShut  [1]{\csname bibitem#1\endcsname}%
\let\auto@bib@innerbib\@empty
\bibitem [{\citenamefont {Bhattacharjee}\ and\ \citenamefont
  {Sigl}(2000)}]{Bhattacharjee:1999mup}%
  \BibitemOpen
  \bibfield  {author} {\bibinfo {author} {\bibfnamefont {P.}~\bibnamefont
  {Bhattacharjee}}\ and\ \bibinfo {author} {\bibfnamefont {G.}~\bibnamefont
  {Sigl}},\ }\href {\doibase 10.1016/S0370-1573(99)00101-5} {\bibfield
  {journal} {\bibinfo  {journal} {Phys. Rept.}\ }\textbf {\bibinfo {volume}
  {327}},\ \bibinfo {pages} {109} (\bibinfo {year} {2000})},\ \Eprint
  {http://arxiv.org/abs/astro-ph/9811011} {arXiv:astro-ph/9811011} \BibitemShut
  {NoStop}%
\bibitem [{\citenamefont {Anchordoqui}(2019)}]{Anchordoqui:2018qom}%
  \BibitemOpen
  \bibfield  {author} {\bibinfo {author} {\bibfnamefont {L.~A.}\ \bibnamefont
  {Anchordoqui}},\ }\href {\doibase 10.1016/j.physrep.2019.01.002} {\bibfield
  {journal} {\bibinfo  {journal} {Phys. Rept.}\ }\textbf {\bibinfo {volume}
  {801}},\ \bibinfo {pages} {1} (\bibinfo {year} {2019})},\ \Eprint
  {http://arxiv.org/abs/1807.09645} {arXiv:1807.09645 [astro-ph.HE]}
  \BibitemShut {NoStop}%
\bibitem [{\citenamefont {Ellis}\ \emph {et~al.}(1992)\citenamefont {Ellis},
  \citenamefont {Gelmini}, \citenamefont {Lopez}, \citenamefont {Nanopoulos},\
  and\ \citenamefont {Sarkar}}]{Ellis:1990nb}%
  \BibitemOpen
  \bibfield  {author} {\bibinfo {author} {\bibfnamefont {J.~R.}\ \bibnamefont
  {Ellis}}, \bibinfo {author} {\bibfnamefont {G.}~\bibnamefont {Gelmini}},
  \bibinfo {author} {\bibfnamefont {J.~L.}\ \bibnamefont {Lopez}}, \bibinfo
  {author} {\bibfnamefont {D.~V.}\ \bibnamefont {Nanopoulos}}, \ and\ \bibinfo
  {author} {\bibfnamefont {S.}~\bibnamefont {Sarkar}},\ }\href {\doibase
  10.1016/0550-3213(92)90438-H} {\bibfield  {journal} {\bibinfo  {journal}
  {Nucl. Phys. B}\ }\textbf {\bibinfo {volume} {373}},\ \bibinfo {pages} {399}
  (\bibinfo {year} {1992})}\BibitemShut {NoStop}%
\bibitem [{\citenamefont {Berezinsky}\ \emph {et~al.}(1997)\citenamefont
  {Berezinsky}, \citenamefont {Kachelrie\ss{}},\ and\ \citenamefont
  {Vilenkin}}]{PhysRevLett.79.4302}%
  \BibitemOpen
  \bibfield  {author} {\bibinfo {author} {\bibfnamefont {V.}~\bibnamefont
  {Berezinsky}}, \bibinfo {author} {\bibfnamefont {M.}~\bibnamefont
  {Kachelrie\ss{}}}, \ and\ \bibinfo {author} {\bibfnamefont {A.}~\bibnamefont
  {Vilenkin}},\ }\href {\doibase 10.1103/PhysRevLett.79.4302} {\bibfield
  {journal} {\bibinfo  {journal} {Phys. Rev. Lett.}\ }\textbf {\bibinfo
  {volume} {79}},\ \bibinfo {pages} {4302} (\bibinfo {year}
  {1997})}\BibitemShut {NoStop}%
\bibitem [{\citenamefont {Chung}\ \emph {et~al.}(1998)\citenamefont {Chung},
  \citenamefont {Kolb},\ and\ \citenamefont {Riotto}}]{Chung:1998zb}%
  \BibitemOpen
  \bibfield  {author} {\bibinfo {author} {\bibfnamefont {D.~J.~H.}\
  \bibnamefont {Chung}}, \bibinfo {author} {\bibfnamefont {E.~W.}\ \bibnamefont
  {Kolb}}, \ and\ \bibinfo {author} {\bibfnamefont {A.}~\bibnamefont
  {Riotto}},\ }\href {\doibase 10.1103/PhysRevD.59.023501} {\bibfield
  {journal} {\bibinfo  {journal} {Phys. Rev. D}\ }\textbf {\bibinfo {volume}
  {59}},\ \bibinfo {pages} {023501} (\bibinfo {year} {1998})},\ \Eprint
  {http://arxiv.org/abs/hep-ph/9802238} {arXiv:hep-ph/9802238} \BibitemShut
  {NoStop}%
\bibitem [{\citenamefont {Kuzmin}\ and\ \citenamefont
  {Tkachev}(1999)}]{Kuzmin:1998kk}%
  \BibitemOpen
  \bibfield  {author} {\bibinfo {author} {\bibfnamefont {V.}~\bibnamefont
  {Kuzmin}}\ and\ \bibinfo {author} {\bibfnamefont {I.}~\bibnamefont
  {Tkachev}},\ }\href {\doibase 10.1103/PhysRevD.59.123006} {\bibfield
  {journal} {\bibinfo  {journal} {Phys. Rev. D}\ }\textbf {\bibinfo {volume}
  {59}},\ \bibinfo {pages} {123006} (\bibinfo {year} {1999})},\ \Eprint
  {http://arxiv.org/abs/hep-ph/9809547} {arXiv:hep-ph/9809547} \BibitemShut
  {NoStop}%
\bibitem [{\citenamefont {Chung}\ \emph {et~al.}(2000)\citenamefont {Chung},
  \citenamefont {Kolb}, \citenamefont {Riotto},\ and\ \citenamefont
  {Tkachev}}]{Chung:1999ve}%
  \BibitemOpen
  \bibfield  {author} {\bibinfo {author} {\bibfnamefont {D.~J.~H.}\
  \bibnamefont {Chung}}, \bibinfo {author} {\bibfnamefont {E.~W.}\ \bibnamefont
  {Kolb}}, \bibinfo {author} {\bibfnamefont {A.}~\bibnamefont {Riotto}}, \ and\
  \bibinfo {author} {\bibfnamefont {I.~I.}\ \bibnamefont {Tkachev}},\ }\href
  {\doibase 10.1103/PhysRevD.62.043508} {\bibfield  {journal} {\bibinfo
  {journal} {Phys. Rev. D}\ }\textbf {\bibinfo {volume} {62}},\ \bibinfo
  {pages} {043508} (\bibinfo {year} {2000})},\ \Eprint
  {http://arxiv.org/abs/hep-ph/9910437} {arXiv:hep-ph/9910437} \BibitemShut
  {NoStop}%
\bibitem [{\citenamefont {Fedderke}\ \emph {et~al.}(2015)\citenamefont
  {Fedderke}, \citenamefont {Kolb},\ and\ \citenamefont
  {Wyman}}]{Fedderke:2014ura}%
  \BibitemOpen
  \bibfield  {author} {\bibinfo {author} {\bibfnamefont {M.~A.}\ \bibnamefont
  {Fedderke}}, \bibinfo {author} {\bibfnamefont {E.~W.}\ \bibnamefont {Kolb}},
  \ and\ \bibinfo {author} {\bibfnamefont {M.}~\bibnamefont {Wyman}},\ }\href
  {\doibase 10.1103/PhysRevD.91.063505} {\bibfield  {journal} {\bibinfo
  {journal} {Phys. Rev. D}\ }\textbf {\bibinfo {volume} {91}},\ \bibinfo
  {pages} {063505} (\bibinfo {year} {2015})},\ \Eprint
  {http://arxiv.org/abs/1409.1584} {arXiv:1409.1584 [astro-ph.CO]} \BibitemShut
  {NoStop}%
\bibitem [{\citenamefont {Garny}\ \emph {et~al.}(2016)\citenamefont {Garny},
  \citenamefont {Sandora},\ and\ \citenamefont {Sloth}}]{Garny:2015sjg}%
  \BibitemOpen
  \bibfield  {author} {\bibinfo {author} {\bibfnamefont {M.}~\bibnamefont
  {Garny}}, \bibinfo {author} {\bibfnamefont {M.}~\bibnamefont {Sandora}}, \
  and\ \bibinfo {author} {\bibfnamefont {M.~S.}\ \bibnamefont {Sloth}},\ }\href
  {\doibase 10.1103/PhysRevLett.116.101302} {\bibfield  {journal} {\bibinfo
  {journal} {Phys. Rev. Lett.}\ }\textbf {\bibinfo {volume} {116}},\ \bibinfo
  {pages} {101302} (\bibinfo {year} {2016})},\ \Eprint
  {http://arxiv.org/abs/1511.03278} {arXiv:1511.03278 [hep-ph]} \BibitemShut
  {NoStop}%
\bibitem [{\citenamefont {Ellis}\ \emph {et~al.}(2016)\citenamefont {Ellis},
  \citenamefont {Garcia}, \citenamefont {Nanopoulos}, \citenamefont {Olive},\
  and\ \citenamefont {Peloso}}]{Ellis:2015jpg}%
  \BibitemOpen
  \bibfield  {author} {\bibinfo {author} {\bibfnamefont {J.}~\bibnamefont
  {Ellis}}, \bibinfo {author} {\bibfnamefont {M.~A.~G.}\ \bibnamefont
  {Garcia}}, \bibinfo {author} {\bibfnamefont {D.~V.}\ \bibnamefont
  {Nanopoulos}}, \bibinfo {author} {\bibfnamefont {K.~A.}\ \bibnamefont
  {Olive}}, \ and\ \bibinfo {author} {\bibfnamefont {M.}~\bibnamefont
  {Peloso}},\ }\href {\doibase 10.1088/1475-7516/2016/03/008} {\bibfield
  {journal} {\bibinfo  {journal} {JCAP}\ }\textbf {\bibinfo {volume} {03}},\
  \bibinfo {pages} {008} (\bibinfo {year} {2016})},\ \Eprint
  {http://arxiv.org/abs/1512.05701} {arXiv:1512.05701 [astro-ph.CO]}
  \BibitemShut {NoStop}%
\bibitem [{\citenamefont {Kolb}\ and\ \citenamefont
  {Long}(2017)}]{Kolb:2017jvz}%
  \BibitemOpen
  \bibfield  {author} {\bibinfo {author} {\bibfnamefont {E.~W.}\ \bibnamefont
  {Kolb}}\ and\ \bibinfo {author} {\bibfnamefont {A.~J.}\ \bibnamefont
  {Long}},\ }\href {\doibase 10.1103/PhysRevD.96.103540} {\bibfield  {journal}
  {\bibinfo  {journal} {Phys. Rev. D}\ }\textbf {\bibinfo {volume} {96}},\
  \bibinfo {pages} {103540} (\bibinfo {year} {2017})},\ \Eprint
  {http://arxiv.org/abs/1708.04293} {arXiv:1708.04293 [astro-ph.CO]}
  \BibitemShut {NoStop}%
\bibitem [{\citenamefont {Dudas}\ \emph {et~al.}(2017)\citenamefont {Dudas},
  \citenamefont {Mambrini},\ and\ \citenamefont {Olive}}]{Dudas:2017rpa}%
  \BibitemOpen
  \bibfield  {author} {\bibinfo {author} {\bibfnamefont {E.}~\bibnamefont
  {Dudas}}, \bibinfo {author} {\bibfnamefont {Y.}~\bibnamefont {Mambrini}}, \
  and\ \bibinfo {author} {\bibfnamefont {K.}~\bibnamefont {Olive}},\ }\href
  {\doibase 10.1103/PhysRevLett.119.051801} {\bibfield  {journal} {\bibinfo
  {journal} {Phys. Rev. Lett.}\ }\textbf {\bibinfo {volume} {119}},\ \bibinfo
  {pages} {051801} (\bibinfo {year} {2017})},\ \Eprint
  {http://arxiv.org/abs/1704.03008} {arXiv:1704.03008 [hep-ph]} \BibitemShut
  {NoStop}%
\bibitem [{\citenamefont {Kaneta}\ \emph {et~al.}(2019)\citenamefont {Kaneta},
  \citenamefont {Mambrini},\ and\ \citenamefont {Olive}}]{Kaneta:2019zgw}%
  \BibitemOpen
  \bibfield  {author} {\bibinfo {author} {\bibfnamefont {K.}~\bibnamefont
  {Kaneta}}, \bibinfo {author} {\bibfnamefont {Y.}~\bibnamefont {Mambrini}}, \
  and\ \bibinfo {author} {\bibfnamefont {K.~A.}\ \bibnamefont {Olive}},\ }\href
  {\doibase 10.1103/PhysRevD.99.063508} {\bibfield  {journal} {\bibinfo
  {journal} {Phys. Rev. D}\ }\textbf {\bibinfo {volume} {99}},\ \bibinfo
  {pages} {063508} (\bibinfo {year} {2019})},\ \Eprint
  {http://arxiv.org/abs/1901.04449} {arXiv:1901.04449 [hep-ph]} \BibitemShut
  {NoStop}%
\bibitem [{\citenamefont {Mambrini}\ and\ \citenamefont
  {Olive}(2021)}]{Mambrini:2021zpp}%
  \BibitemOpen
  \bibfield  {author} {\bibinfo {author} {\bibfnamefont {Y.}~\bibnamefont
  {Mambrini}}\ and\ \bibinfo {author} {\bibfnamefont {K.~A.}\ \bibnamefont
  {Olive}},\ }\href {\doibase 10.1103/PhysRevD.103.115009} {\bibfield
  {journal} {\bibinfo  {journal} {Phys. Rev. D}\ }\textbf {\bibinfo {volume}
  {103}},\ \bibinfo {pages} {115009} (\bibinfo {year} {2021})},\ \Eprint
  {http://arxiv.org/abs/2102.06214} {arXiv:2102.06214 [hep-ph]} \BibitemShut
  {NoStop}%
\bibitem [{\citenamefont {Ade}\ \emph {et~al.}(2015)\citenamefont {Ade} \emph
  {et~al.}}]{BICEP2:2015nss}%
  \BibitemOpen
  \bibfield  {author} {\bibinfo {author} {\bibfnamefont {P.~A.~R.}\
  \bibnamefont {Ade}} \emph {et~al.} (\bibinfo {collaboration} {BICEP2,
  Planck}),\ }\href {\doibase 10.1103/PhysRevLett.114.101301} {\bibfield
  {journal} {\bibinfo  {journal} {Phys. Rev. Lett.}\ }\textbf {\bibinfo
  {volume} {114}},\ \bibinfo {pages} {101301} (\bibinfo {year} {2015})},\
  \Eprint {http://arxiv.org/abs/1502.00612} {arXiv:1502.00612 [astro-ph.CO]}
  \BibitemShut {NoStop}%
\bibitem [{\citenamefont {Ade}\ \emph {et~al.}(2016)\citenamefont {Ade} \emph
  {et~al.}}]{Planck:2015sxf}%
  \BibitemOpen
  \bibfield  {author} {\bibinfo {author} {\bibfnamefont {P.~A.~R.}\
  \bibnamefont {Ade}} \emph {et~al.} (\bibinfo {collaboration} {Planck}),\
  }\href {\doibase 10.1051/0004-6361/201525898} {\bibfield  {journal} {\bibinfo
   {journal} {Astron. Astrophys.}\ }\textbf {\bibinfo {volume} {594}},\
  \bibinfo {pages} {A20} (\bibinfo {year} {2016})},\ \Eprint
  {http://arxiv.org/abs/1502.02114} {arXiv:1502.02114 [astro-ph.CO]}
  \BibitemShut {NoStop}%
\bibitem [{\citenamefont {Zyla}\ \emph {et~al.}(2020)\citenamefont {Zyla} \emph
  {et~al.}}]{10.1093/ptep/ptaa104}%
  \BibitemOpen
  \bibfield  {author} {\bibinfo {author} {\bibfnamefont {P.~A.}\ \bibnamefont
  {Zyla}} \emph {et~al.} (\bibinfo {collaboration} {Particle Data Group}),\
  }\href {\doibase 10.1093/ptep/ptaa104} {\bibfield  {journal} {\bibinfo
  {journal} {PTEP}\ }\textbf {\bibinfo {volume} {2020}} (\bibinfo {year}
  {2020}),\ 10.1093/ptep/ptaa104},\ \bibinfo {note} {083C01},\ \Eprint
  {http://arxiv.org/abs/https://academic.oup.com/ptep/article-pdf/2020/8/083C01/34673722/ptaa104.pdf}
  {https://academic.oup.com/ptep/article-pdf/2020/8/083C01/34673722/ptaa104.pdf}
  \BibitemShut {NoStop}%
\bibitem [{\citenamefont {'t~Hooft}(1980)}]{tHooft:1979rat}%
  \BibitemOpen
  \bibfield  {author} {\bibinfo {author} {\bibfnamefont {G.}~\bibnamefont
  {'t~Hooft}},\ }\href {\doibase 10.1007/978-1-4684-7571-5_9} {\bibfield
  {journal} {\bibinfo  {journal} {NATO Sci. Ser. B}\ }\textbf {\bibinfo
  {volume} {59}},\ \bibinfo {pages} {135} (\bibinfo {year} {1980})}\BibitemShut
  {NoStop}%
\bibitem [{\citenamefont {Marrod\'an~Undagoitia}\ and\ \citenamefont
  {Rauch}(2016)}]{MarrodanUndagoitia:2015veg}%
  \BibitemOpen
  \bibfield  {author} {\bibinfo {author} {\bibfnamefont {T.}~\bibnamefont
  {Marrod\'an~Undagoitia}}\ and\ \bibinfo {author} {\bibfnamefont
  {L.}~\bibnamefont {Rauch}},\ }\href {\doibase 10.1088/0954-3899/43/1/013001}
  {\bibfield  {journal} {\bibinfo  {journal} {J. Phys. G}\ }\textbf {\bibinfo
  {volume} {43}},\ \bibinfo {pages} {013001} (\bibinfo {year} {2016})},\
  \Eprint {http://arxiv.org/abs/1509.08767} {arXiv:1509.08767
  [physics.ins-det]} \BibitemShut {NoStop}%
\bibitem [{\citenamefont {Rappoccio}(2019)}]{Rappoccio:2018qxp}%
  \BibitemOpen
  \bibfield  {author} {\bibinfo {author} {\bibfnamefont {S.}~\bibnamefont
  {Rappoccio}},\ }\href {\doibase 10.1016/j.revip.2018.100027} {\bibfield
  {journal} {\bibinfo  {journal} {Rev. Phys.}\ }\textbf {\bibinfo {volume}
  {4}},\ \bibinfo {pages} {100027} (\bibinfo {year} {2019})},\ \Eprint
  {http://arxiv.org/abs/1810.10579} {arXiv:1810.10579 [hep-ex]} \BibitemShut
  {NoStop}%
\bibitem [{\citenamefont {Gaskins}(2016)}]{Gaskins:2016cha}%
  \BibitemOpen
  \bibfield  {author} {\bibinfo {author} {\bibfnamefont {J.~M.}\ \bibnamefont
  {Gaskins}},\ }\href {\doibase 10.1080/00107514.2016.1175160} {\bibfield
  {journal} {\bibinfo  {journal} {Contemp. Phys.}\ }\textbf {\bibinfo {volume}
  {57}},\ \bibinfo {pages} {496} (\bibinfo {year} {2016})},\ \Eprint
  {http://arxiv.org/abs/1604.00014} {arXiv:1604.00014 [astro-ph.HE]}
  \BibitemShut {NoStop}%
\bibitem [{\citenamefont {Buttazzo}\ \emph {et~al.}(2013)\citenamefont
  {Buttazzo}, \citenamefont {Degrassi}, \citenamefont {Giardino}, \citenamefont
  {Giudice}, \citenamefont {Sala}, \citenamefont {Salvio},\ and\ \citenamefont
  {Strumia}}]{Buttazzo:2013uya}%
  \BibitemOpen
  \bibfield  {author} {\bibinfo {author} {\bibfnamefont {D.}~\bibnamefont
  {Buttazzo}}, \bibinfo {author} {\bibfnamefont {G.}~\bibnamefont {Degrassi}},
  \bibinfo {author} {\bibfnamefont {P.~P.}\ \bibnamefont {Giardino}}, \bibinfo
  {author} {\bibfnamefont {G.~F.}\ \bibnamefont {Giudice}}, \bibinfo {author}
  {\bibfnamefont {F.}~\bibnamefont {Sala}}, \bibinfo {author} {\bibfnamefont
  {A.}~\bibnamefont {Salvio}}, \ and\ \bibinfo {author} {\bibfnamefont
  {A.}~\bibnamefont {Strumia}},\ }\href {\doibase 10.1007/JHEP12(2013)089}
  {\bibfield  {journal} {\bibinfo  {journal} {JHEP}\ }\textbf {\bibinfo
  {volume} {12}},\ \bibinfo {pages} {089} (\bibinfo {year} {2013})},\ \Eprint
  {http://arxiv.org/abs/1307.3536} {arXiv:1307.3536 [hep-ph]} \BibitemShut
  {NoStop}%
\bibitem [{\citenamefont {Alekhin}\ \emph {et~al.}(2012)\citenamefont
  {Alekhin}, \citenamefont {Djouadi},\ and\ \citenamefont
  {Moch}}]{Alekhin:2012py}%
  \BibitemOpen
  \bibfield  {author} {\bibinfo {author} {\bibfnamefont {S.}~\bibnamefont
  {Alekhin}}, \bibinfo {author} {\bibfnamefont {A.}~\bibnamefont {Djouadi}}, \
  and\ \bibinfo {author} {\bibfnamefont {S.}~\bibnamefont {Moch}},\ }\href
  {\doibase 10.1016/j.physletb.2012.08.024} {\bibfield  {journal} {\bibinfo
  {journal} {Phys. Lett. B}\ }\textbf {\bibinfo {volume} {716}},\ \bibinfo
  {pages} {214} (\bibinfo {year} {2012})},\ \Eprint
  {http://arxiv.org/abs/1207.0980} {arXiv:1207.0980 [hep-ph]} \BibitemShut
  {NoStop}%
\bibitem [{\citenamefont {Bednyakov}\ \emph {et~al.}(2015)\citenamefont
  {Bednyakov}, \citenamefont {Kniehl}, \citenamefont {Pikelner},\ and\
  \citenamefont {Veretin}}]{Bednyakov:2015sca}%
  \BibitemOpen
  \bibfield  {author} {\bibinfo {author} {\bibfnamefont {A.~V.}\ \bibnamefont
  {Bednyakov}}, \bibinfo {author} {\bibfnamefont {B.~A.}\ \bibnamefont
  {Kniehl}}, \bibinfo {author} {\bibfnamefont {A.~F.}\ \bibnamefont
  {Pikelner}}, \ and\ \bibinfo {author} {\bibfnamefont {O.~L.}\ \bibnamefont
  {Veretin}},\ }\href {\doibase 10.1103/PhysRevLett.115.201802} {\bibfield
  {journal} {\bibinfo  {journal} {Phys. Rev. Lett.}\ }\textbf {\bibinfo
  {volume} {115}},\ \bibinfo {pages} {201802} (\bibinfo {year} {2015})},\
  \Eprint {http://arxiv.org/abs/1507.08833} {arXiv:1507.08833 [hep-ph]}
  \BibitemShut {NoStop}%
\bibitem [{\citenamefont {Damour}\ and\ \citenamefont
  {Donoghue}(2008)}]{Damour:2007uv}%
  \BibitemOpen
  \bibfield  {author} {\bibinfo {author} {\bibfnamefont {T.}~\bibnamefont
  {Damour}}\ and\ \bibinfo {author} {\bibfnamefont {J.~F.}\ \bibnamefont
  {Donoghue}},\ }\href {\doibase 10.1103/PhysRevD.78.014014} {\bibfield
  {journal} {\bibinfo  {journal} {Phys. Rev. D}\ }\textbf {\bibinfo {volume}
  {78}},\ \bibinfo {pages} {014014} (\bibinfo {year} {2008})},\ \Eprint
  {http://arxiv.org/abs/0712.2968} {arXiv:0712.2968 [hep-ph]} \BibitemShut
  {NoStop}%
\bibitem [{\citenamefont {de~Vega}\ and\ \citenamefont
  {Sanchez}(2003)}]{deVega:2003hh}%
  \BibitemOpen
  \bibfield  {author} {\bibinfo {author} {\bibfnamefont {H.~J.}\ \bibnamefont
  {de~Vega}}\ and\ \bibinfo {author} {\bibfnamefont {N.~G.}\ \bibnamefont
  {Sanchez}},\ }\href {\doibase 10.1103/PhysRevD.67.125019} {\bibfield
  {journal} {\bibinfo  {journal} {Phys. Rev. D}\ }\textbf {\bibinfo {volume}
  {67}},\ \bibinfo {pages} {125019} (\bibinfo {year} {2003})}\BibitemShut
  {NoStop}%
\bibitem [{\citenamefont {Abreu}\ \emph
  {et~al.}(2022{\natexlab{a}})\citenamefont {Abreu} \emph
  {et~al.}}]{PierreAuger:2022ibr}%
  \BibitemOpen
  \bibfield  {author} {\bibinfo {author} {\bibfnamefont {P.}~\bibnamefont
  {Abreu}} \emph {et~al.} (\bibinfo {collaboration} {Pierre Auger}),\
  }\href@noop {} {\  (\bibinfo {year} {2022}{\natexlab{a}})},\ \Eprint
  {http://arxiv.org/abs/2208.02353} {arXiv:2208.02353 [astro-ph.HE]}
  \BibitemShut {NoStop}%
\bibitem [{\citenamefont {Belavin}\ \emph {et~al.}(1975)\citenamefont
  {Belavin}, \citenamefont {Polyakov}, \citenamefont {Schwartz},\ and\
  \citenamefont {Tyupkin}}]{Belavin:1975fg}%
  \BibitemOpen
  \bibfield  {author} {\bibinfo {author} {\bibfnamefont {A.~A.}\ \bibnamefont
  {Belavin}}, \bibinfo {author} {\bibfnamefont {A.~M.}\ \bibnamefont
  {Polyakov}}, \bibinfo {author} {\bibfnamefont {A.~S.}\ \bibnamefont
  {Schwartz}}, \ and\ \bibinfo {author} {\bibfnamefont {Y.~S.}\ \bibnamefont
  {Tyupkin}},\ }\href {\doibase 10.1016/0370-2693(75)90163-X} {\bibfield
  {journal} {\bibinfo  {journal} {Phys. Lett. B}\ }\textbf {\bibinfo {volume}
  {59}},\ \bibinfo {pages} {85} (\bibinfo {year} {1975})}\BibitemShut {NoStop}%
\bibitem [{\citenamefont {Coleman}(1979)}]{Coleman:1978ae}%
  \BibitemOpen
  \bibfield  {author} {\bibinfo {author} {\bibfnamefont {S.~R.}\ \bibnamefont
  {Coleman}},\ }\href@noop {} {\bibfield  {journal} {\bibinfo  {journal}
  {Subnucl. Ser.}\ }\textbf {\bibinfo {volume} {15}},\ \bibinfo {pages} {805}
  (\bibinfo {year} {1979})}\BibitemShut {NoStop}%
\bibitem [{\citenamefont {Vainshtein}\ \emph {et~al.}(1982)\citenamefont
  {Vainshtein}, \citenamefont {Zakharov}, \citenamefont {Novikov},\ and\
  \citenamefont {Shifman}}]{Vainshtein:1981wh}%
  \BibitemOpen
  \bibfield  {author} {\bibinfo {author} {\bibfnamefont {A.~I.}\ \bibnamefont
  {Vainshtein}}, \bibinfo {author} {\bibfnamefont {V.~I.}\ \bibnamefont
  {Zakharov}}, \bibinfo {author} {\bibfnamefont {V.~A.}\ \bibnamefont
  {Novikov}}, \ and\ \bibinfo {author} {\bibfnamefont {M.~A.}\ \bibnamefont
  {Shifman}},\ }\href {\doibase 10.1070/PU1982v025n04ABEH004533} {\bibfield
  {journal} {\bibinfo  {journal} {Sov. Phys. Usp.}\ }\textbf {\bibinfo {volume}
  {25}},\ \bibinfo {pages} {195} (\bibinfo {year} {1982})}\BibitemShut
  {NoStop}%
\bibitem [{\citenamefont {Kuzmin}\ and\ \citenamefont
  {Rubakov}(1998)}]{Kuzmin:1997jua}%
  \BibitemOpen
  \bibfield  {author} {\bibinfo {author} {\bibfnamefont {V.~A.}\ \bibnamefont
  {Kuzmin}}\ and\ \bibinfo {author} {\bibfnamefont {V.~A.}\ \bibnamefont
  {Rubakov}},\ }\href@noop {} {\bibfield  {journal} {\bibinfo  {journal} {Phys.
  Atom. Nucl.}\ }\textbf {\bibinfo {volume} {61}},\ \bibinfo {pages} {1028}
  (\bibinfo {year} {1998})},\ \Eprint {http://arxiv.org/abs/astro-ph/9709187}
  {arXiv:astro-ph/9709187} \BibitemShut {NoStop}%
\bibitem [{\citenamefont {'t~Hooft}(1976{\natexlab{a}})}]{tHooft:1976rip}%
  \BibitemOpen
  \bibfield  {author} {\bibinfo {author} {\bibfnamefont {G.}~\bibnamefont
  {'t~Hooft}},\ }\href {\doibase 10.1103/PhysRevLett.37.8} {\bibfield
  {journal} {\bibinfo  {journal} {Phys. Rev. Lett.}\ }\textbf {\bibinfo
  {volume} {37}},\ \bibinfo {pages} {8} (\bibinfo {year}
  {1976}{\natexlab{a}})}\BibitemShut {NoStop}%
\bibitem [{\citenamefont {Aloisio}\ \emph {et~al.}(2004)\citenamefont
  {Aloisio}, \citenamefont {Berezinsky},\ and\ \citenamefont
  {Kachelriess}}]{Aloisio:2003xj}%
  \BibitemOpen
  \bibfield  {author} {\bibinfo {author} {\bibfnamefont {R.}~\bibnamefont
  {Aloisio}}, \bibinfo {author} {\bibfnamefont {V.}~\bibnamefont {Berezinsky}},
  \ and\ \bibinfo {author} {\bibfnamefont {M.}~\bibnamefont {Kachelriess}},\
  }\href {\doibase 10.1103/PhysRevD.69.094023} {\bibfield  {journal} {\bibinfo
  {journal} {Phys. Rev. D}\ }\textbf {\bibinfo {volume} {69}},\ \bibinfo
  {pages} {094023} (\bibinfo {year} {2004})},\ \Eprint
  {http://arxiv.org/abs/hep-ph/0307279} {arXiv:hep-ph/0307279} \BibitemShut
  {NoStop}%
\bibitem [{\citenamefont {Sarkar}\ and\ \citenamefont
  {Toldra}(2002)}]{Sarkar:2001se}%
  \BibitemOpen
  \bibfield  {author} {\bibinfo {author} {\bibfnamefont {S.}~\bibnamefont
  {Sarkar}}\ and\ \bibinfo {author} {\bibfnamefont {R.}~\bibnamefont
  {Toldra}},\ }\href {\doibase 10.1016/S0550-3213(01)00565-X} {\bibfield
  {journal} {\bibinfo  {journal} {Nucl. Phys. B}\ }\textbf {\bibinfo {volume}
  {621}},\ \bibinfo {pages} {495} (\bibinfo {year} {2002})},\ \Eprint
  {http://arxiv.org/abs/hep-ph/0108098} {arXiv:hep-ph/0108098} \BibitemShut
  {NoStop}%
\bibitem [{\citenamefont {Barbot}\ and\ \citenamefont
  {Drees}(2003)}]{Barbot:2002gt}%
  \BibitemOpen
  \bibfield  {author} {\bibinfo {author} {\bibfnamefont {C.}~\bibnamefont
  {Barbot}}\ and\ \bibinfo {author} {\bibfnamefont {M.}~\bibnamefont {Drees}},\
  }\href {\doibase 10.1016/S0927-6505(03)00134-8} {\bibfield  {journal}
  {\bibinfo  {journal} {Astropart. Phys.}\ }\textbf {\bibinfo {volume} {20}},\
  \bibinfo {pages} {5} (\bibinfo {year} {2003})},\ \Eprint
  {http://arxiv.org/abs/hep-ph/0211406} {arXiv:hep-ph/0211406} \BibitemShut
  {NoStop}%
\bibitem [{\citenamefont {Kachelriess}\ \emph {et~al.}(2018)\citenamefont
  {Kachelriess}, \citenamefont {Kalashev},\ and\ \citenamefont
  {Kuznetsov}}]{Kachelriess:2018rty}%
  \BibitemOpen
  \bibfield  {author} {\bibinfo {author} {\bibfnamefont {M.}~\bibnamefont
  {Kachelriess}}, \bibinfo {author} {\bibfnamefont {O.~E.}\ \bibnamefont
  {Kalashev}}, \ and\ \bibinfo {author} {\bibfnamefont {M.~Y.}\ \bibnamefont
  {Kuznetsov}},\ }\href {\doibase 10.1103/PhysRevD.98.083016} {\bibfield
  {journal} {\bibinfo  {journal} {Phys. Rev. D}\ }\textbf {\bibinfo {volume}
  {98}},\ \bibinfo {pages} {083016} (\bibinfo {year} {2018})},\ \Eprint
  {http://arxiv.org/abs/1805.04500} {arXiv:1805.04500 [astro-ph.HE]}
  \BibitemShut {NoStop}%
\bibitem [{\citenamefont {Alcantara}\ \emph {et~al.}(2019)\citenamefont
  {Alcantara}, \citenamefont {Anchordoqui},\ and\ \citenamefont
  {Soriano}}]{Alcantara:2019sco}%
  \BibitemOpen
  \bibfield  {author} {\bibinfo {author} {\bibfnamefont {E.}~\bibnamefont
  {Alcantara}}, \bibinfo {author} {\bibfnamefont {L.~A.}\ \bibnamefont
  {Anchordoqui}}, \ and\ \bibinfo {author} {\bibfnamefont {J.~F.}\ \bibnamefont
  {Soriano}},\ }\href {\doibase 10.1103/PhysRevD.99.103016} {\bibfield
  {journal} {\bibinfo  {journal} {Phys. Rev. D}\ }\textbf {\bibinfo {volume}
  {99}},\ \bibinfo {pages} {103016} (\bibinfo {year} {2019})},\ \Eprint
  {http://arxiv.org/abs/1903.05429} {arXiv:1903.05429 [hep-ph]} \BibitemShut
  {NoStop}%
\bibitem [{\citenamefont {Barbot}\ \emph {et~al.}(2003)\citenamefont {Barbot},
  \citenamefont {Drees}, \citenamefont {Halzen},\ and\ \citenamefont
  {Hooper}}]{Barbot:2002et}%
  \BibitemOpen
  \bibfield  {author} {\bibinfo {author} {\bibfnamefont {C.}~\bibnamefont
  {Barbot}}, \bibinfo {author} {\bibfnamefont {M.}~\bibnamefont {Drees}},
  \bibinfo {author} {\bibfnamefont {F.}~\bibnamefont {Halzen}}, \ and\ \bibinfo
  {author} {\bibfnamefont {D.}~\bibnamefont {Hooper}},\ }\href {\doibase
  10.1016/S0370-2693(03)00642-7} {\bibfield  {journal} {\bibinfo  {journal}
  {Phys. Lett. B}\ }\textbf {\bibinfo {volume} {563}},\ \bibinfo {pages} {132}
  (\bibinfo {year} {2003})},\ \Eprint {http://arxiv.org/abs/hep-ph/0207133}
  {arXiv:hep-ph/0207133} \BibitemShut {NoStop}%
\bibitem [{\citenamefont {Navarro}\ \emph {et~al.}(1996)\citenamefont
  {Navarro}, \citenamefont {Frenk},\ and\ \citenamefont
  {White}}]{Navarro:1995iw}%
  \BibitemOpen
  \bibfield  {author} {\bibinfo {author} {\bibfnamefont {J.~F.}\ \bibnamefont
  {Navarro}}, \bibinfo {author} {\bibfnamefont {C.~S.}\ \bibnamefont {Frenk}},
  \ and\ \bibinfo {author} {\bibfnamefont {S.~D.~M.}\ \bibnamefont {White}},\
  }\href {\doibase 10.1086/177173} {\bibfield  {journal} {\bibinfo  {journal}
  {Astrophys. J.}\ }\textbf {\bibinfo {volume} {462}},\ \bibinfo {pages} {563}
  (\bibinfo {year} {1996})},\ \Eprint {http://arxiv.org/abs/astro-ph/9508025}
  {arXiv:astro-ph/9508025} \BibitemShut {NoStop}%
\bibitem [{\citenamefont {Einasto}(1965)}]{Einasto:1965czb}%
  \BibitemOpen
  \bibfield  {author} {\bibinfo {author} {\bibfnamefont {J.}~\bibnamefont
  {Einasto}},\ }\href@noop {} {\bibfield  {journal} {\bibinfo  {journal} {Trudy
  Astrofizicheskogo Instituta Alma-Ata}\ }\textbf {\bibinfo {volume} {5}},\
  \bibinfo {pages} {87} (\bibinfo {year} {1965})}\BibitemShut {NoStop}%
\bibitem [{\citenamefont {Burkert}(1995)}]{Burkert:1995yz}%
  \BibitemOpen
  \bibfield  {author} {\bibinfo {author} {\bibfnamefont {A.}~\bibnamefont
  {Burkert}},\ }\href {\doibase 10.1086/309560} {\bibfield  {journal} {\bibinfo
   {journal} {Astrophys. J. Lett.}\ }\textbf {\bibinfo {volume} {447}},\
  \bibinfo {pages} {L25} (\bibinfo {year} {1995})},\ \Eprint
  {http://arxiv.org/abs/astro-ph/9504041} {arXiv:astro-ph/9504041} \BibitemShut
  {NoStop}%
\bibitem [{\citenamefont {Moore}\ \emph {et~al.}(1999)\citenamefont {Moore},
  \citenamefont {Ghigna}, \citenamefont {Governato}, \citenamefont {Lake},
  \citenamefont {Quinn}, \citenamefont {Stadel},\ and\ \citenamefont
  {Tozzi}}]{Moore:1999nt}%
  \BibitemOpen
  \bibfield  {author} {\bibinfo {author} {\bibfnamefont {B.}~\bibnamefont
  {Moore}}, \bibinfo {author} {\bibfnamefont {S.}~\bibnamefont {Ghigna}},
  \bibinfo {author} {\bibfnamefont {F.}~\bibnamefont {Governato}}, \bibinfo
  {author} {\bibfnamefont {G.}~\bibnamefont {Lake}}, \bibinfo {author}
  {\bibfnamefont {T.~R.}\ \bibnamefont {Quinn}}, \bibinfo {author}
  {\bibfnamefont {J.}~\bibnamefont {Stadel}}, \ and\ \bibinfo {author}
  {\bibfnamefont {P.}~\bibnamefont {Tozzi}},\ }\href {\doibase 10.1086/312287}
  {\bibfield  {journal} {\bibinfo  {journal} {Astrophys. J. Lett.}\ }\textbf
  {\bibinfo {volume} {524}},\ \bibinfo {pages} {L19} (\bibinfo {year}
  {1999})},\ \Eprint {http://arxiv.org/abs/astro-ph/9907411}
  {arXiv:astro-ph/9907411} \BibitemShut {NoStop}%
\bibitem [{\citenamefont {Aab}\ \emph {et~al.}(2015)\citenamefont {Aab} \emph
  {et~al.}}]{PierreAuger:2015eyc}%
  \BibitemOpen
  \bibfield  {author} {\bibinfo {author} {\bibfnamefont {A.}~\bibnamefont
  {Aab}} \emph {et~al.} (\bibinfo {collaboration} {Pierre Auger}),\ }\href
  {\doibase 10.1016/j.nima.2015.06.058} {\bibfield  {journal} {\bibinfo
  {journal} {Nucl. Instrum. Meth. A}\ }\textbf {\bibinfo {volume} {798}},\
  \bibinfo {pages} {172} (\bibinfo {year} {2015})},\ \Eprint
  {http://arxiv.org/abs/1502.01323} {arXiv:1502.01323 [astro-ph.IM]}
  \BibitemShut {NoStop}%
\bibitem [{\citenamefont {Abreu}\ \emph
  {et~al.}(2022{\natexlab{b}})\citenamefont {Abreu} \emph
  {et~al.}}]{PierreAuger:2022uwd}%
  \BibitemOpen
  \bibfield  {author} {\bibinfo {author} {\bibfnamefont {P.}~\bibnamefont
  {Abreu}} \emph {et~al.} (\bibinfo {collaboration} {Pierre Auger}),\ }\href
  {\doibase 10.3847/1538-4357/ac7393} {\bibfield  {journal} {\bibinfo
  {journal} {Astrophys. J.}\ }\textbf {\bibinfo {volume} {933}},\ \bibinfo
  {pages} {125} (\bibinfo {year} {2022}{\natexlab{b}})},\ \Eprint
  {http://arxiv.org/abs/2205.14864} {arXiv:2205.14864 [astro-ph.HE]}
  \BibitemShut {NoStop}%
\bibitem [{\citenamefont {Savina}(2021)}]{Savina:2021ufs}%
  \BibitemOpen
  \bibfield  {author} {\bibinfo {author} {\bibfnamefont {P.}~\bibnamefont
  {Savina}} (\bibinfo {collaboration} {Pierre Auger}),\ }\href {\doibase
  10.22323/1.395.0373} {\bibfield  {journal} {\bibinfo  {journal} {Proc. 37th
  International Cosmic Ray Conference (Berlin, Germany)}\ }\textbf {\bibinfo
  {volume} {PoS (ICRC2021)}},\ \bibinfo {pages} {373} (\bibinfo {year}
  {2021})}\BibitemShut {NoStop}%
\bibitem [{\citenamefont {Abreu}\ \emph
  {et~al.}(2022{\natexlab{c}})\citenamefont {Abreu} \emph
  {et~al.}}]{PierreAuger:2022aty}%
  \BibitemOpen
  \bibfield  {author} {\bibinfo {author} {\bibfnamefont {P.}~\bibnamefont
  {Abreu}} \emph {et~al.} (\bibinfo {collaboration} {Pierre Auger}),\
  }\href@noop {} {\  (\bibinfo {year} {2022}{\natexlab{c}})},\ \Eprint
  {http://arxiv.org/abs/2209.05926} {arXiv:2209.05926 [astro-ph.HE]}
  \BibitemShut {NoStop}%
\bibitem [{\citenamefont {Aab}\ \emph {et~al.}(2020)\citenamefont {Aab} \emph
  {et~al.}}]{PierreAuger:2020qqz}%
  \BibitemOpen
  \bibfield  {author} {\bibinfo {author} {\bibfnamefont {A.}~\bibnamefont
  {Aab}} \emph {et~al.} (\bibinfo {collaboration} {Pierre Auger}),\ }\href
  {\doibase 10.1103/PhysRevD.102.062005} {\bibfield  {journal} {\bibinfo
  {journal} {Phys. Rev. D}\ }\textbf {\bibinfo {volume} {102}},\ \bibinfo
  {pages} {062005} (\bibinfo {year} {2020})},\ \Eprint
  {http://arxiv.org/abs/2008.06486} {arXiv:2008.06486 [astro-ph.HE]}
  \BibitemShut {NoStop}%
\bibitem [{\citenamefont {Anchordoqui}\ \emph {et~al.}(2021)\citenamefont
  {Anchordoqui} \emph {et~al.}}]{Anchordoqui:2021crl}%
  \BibitemOpen
  \bibfield  {author} {\bibinfo {author} {\bibfnamefont {L.~A.}\ \bibnamefont
  {Anchordoqui}} \emph {et~al.},\ }\href {\doibase
  10.1016/j.astropartphys.2021.102614} {\bibfield  {journal} {\bibinfo
  {journal} {Astropart. Phys.}\ }\textbf {\bibinfo {volume} {132}},\ \bibinfo
  {pages} {102614} (\bibinfo {year} {2021})},\ \Eprint
  {http://arxiv.org/abs/2105.12895} {arXiv:2105.12895 [hep-ph]} \BibitemShut
  {NoStop}%
\bibitem [{\citenamefont {Aab}\ \emph {et~al.}(2019)\citenamefont {Aab} \emph
  {et~al.}}]{PierreAuger:2019ens}%
  \BibitemOpen
  \bibfield  {author} {\bibinfo {author} {\bibfnamefont {A.}~\bibnamefont
  {Aab}} \emph {et~al.} (\bibinfo {collaboration} {Pierre Auger}),\ }\href
  {\doibase 10.1088/1475-7516/2019/10/022} {\bibfield  {journal} {\bibinfo
  {journal} {JCAP}\ }\textbf {\bibinfo {volume} {10}},\ \bibinfo {pages} {022}
  (\bibinfo {year} {2019})},\ \Eprint {http://arxiv.org/abs/1906.07422}
  {arXiv:1906.07422 [astro-ph.HE]} \BibitemShut {NoStop}%
\bibitem [{\citenamefont {'t~Hooft}(1976{\natexlab{b}})}]{tHooft:1976snw}%
  \BibitemOpen
  \bibfield  {author} {\bibinfo {author} {\bibfnamefont {G.}~\bibnamefont
  {'t~Hooft}},\ }\href {\doibase 10.1103/PhysRevD.14.3432} {\bibfield
  {journal} {\bibinfo  {journal} {Phys. Rev. D}\ }\textbf {\bibinfo {volume}
  {14}},\ \bibinfo {pages} {3432} (\bibinfo {year} {1976}{\natexlab{b}})},\
  \bibinfo {note} {[Erratum: Phys.Rev.D 18, 2199 (1978)]}\BibitemShut {NoStop}%
\bibitem [{\citenamefont {Giudice}\ \emph {et~al.}(2001)\citenamefont
  {Giudice}, \citenamefont {Kolb},\ and\ \citenamefont
  {Riotto}}]{Giudice:2000ex}%
  \BibitemOpen
  \bibfield  {author} {\bibinfo {author} {\bibfnamefont {G.~F.}\ \bibnamefont
  {Giudice}}, \bibinfo {author} {\bibfnamefont {E.~W.}\ \bibnamefont {Kolb}}, \
  and\ \bibinfo {author} {\bibfnamefont {A.}~\bibnamefont {Riotto}},\ }\href
  {\doibase 10.1103/PhysRevD.64.023508} {\bibfield  {journal} {\bibinfo
  {journal} {Phys. Rev. D}\ }\textbf {\bibinfo {volume} {64}},\ \bibinfo
  {pages} {023508} (\bibinfo {year} {2001})},\ \Eprint
  {http://arxiv.org/abs/hep-ph/0005123} {arXiv:hep-ph/0005123} \BibitemShut
  {NoStop}%
\bibitem [{\citenamefont {Kolb}\ and\ \citenamefont
  {Turner}(1990)}]{Kolb:1990vq}%
  \BibitemOpen
  \bibfield  {author} {\bibinfo {author} {\bibfnamefont {E.~W.}\ \bibnamefont
  {Kolb}}\ and\ \bibinfo {author} {\bibfnamefont {M.~S.}\ \bibnamefont
  {Turner}},\ }\href {\doibase 10.1201/9780429492860} {\emph {\bibinfo {title}
  {{The Early Universe}}}},\ Vol.~\bibinfo {volume} {69}\ (\bibinfo {year}
  {1990})\BibitemShut {NoStop}%
\bibitem [{\citenamefont {Chung}\ \emph {et~al.}(1999)\citenamefont {Chung},
  \citenamefont {Kolb},\ and\ \citenamefont {Riotto}}]{Chung:1998rq}%
  \BibitemOpen
  \bibfield  {author} {\bibinfo {author} {\bibfnamefont {D.~J.~H.}\
  \bibnamefont {Chung}}, \bibinfo {author} {\bibfnamefont {E.~W.}\ \bibnamefont
  {Kolb}}, \ and\ \bibinfo {author} {\bibfnamefont {A.}~\bibnamefont
  {Riotto}},\ }\href {\doibase 10.1103/PhysRevD.60.063504} {\bibfield
  {journal} {\bibinfo  {journal} {Phys. Rev. D}\ }\textbf {\bibinfo {volume}
  {60}},\ \bibinfo {pages} {063504} (\bibinfo {year} {1999})},\ \Eprint
  {http://arxiv.org/abs/hep-ph/9809453} {arXiv:hep-ph/9809453} \BibitemShut
  {NoStop}%
\bibitem [{\citenamefont {Garny}\ \emph {et~al.}(2018)\citenamefont {Garny},
  \citenamefont {Palessandro}, \citenamefont {Sandora},\ and\ \citenamefont
  {Sloth}}]{Garny:2017kha}%
  \BibitemOpen
  \bibfield  {author} {\bibinfo {author} {\bibfnamefont {M.}~\bibnamefont
  {Garny}}, \bibinfo {author} {\bibfnamefont {A.}~\bibnamefont {Palessandro}},
  \bibinfo {author} {\bibfnamefont {M.}~\bibnamefont {Sandora}}, \ and\
  \bibinfo {author} {\bibfnamefont {M.~S.}\ \bibnamefont {Sloth}},\ }\href
  {\doibase 10.1088/1475-7516/2018/02/027} {\bibfield  {journal} {\bibinfo
  {journal} {JCAP}\ }\textbf {\bibinfo {volume} {02}},\ \bibinfo {pages} {027}
  (\bibinfo {year} {2018})},\ \Eprint {http://arxiv.org/abs/1709.09688}
  {arXiv:1709.09688 [hep-ph]} \BibitemShut {NoStop}%
\bibitem [{\citenamefont {Abazajian}\ \emph {et~al.}(2016)\citenamefont
  {Abazajian} \emph {et~al.}}]{CMB-S4:2016ple}%
  \BibitemOpen
  \bibfield  {author} {\bibinfo {author} {\bibfnamefont {K.~N.}\ \bibnamefont
  {Abazajian}} \emph {et~al.} (\bibinfo {collaboration} {CMB-S4}),\ }\href@noop
  {} {\  (\bibinfo {year} {2016})},\ \Eprint {http://arxiv.org/abs/1610.02743}
  {arXiv:1610.02743 [astro-ph.CO]} \BibitemShut {NoStop}%
\bibitem [{\citenamefont {Ishino}(2016)}]{Ishino:2016wxl}%
  \BibitemOpen
  \bibfield  {author} {\bibinfo {author} {\bibfnamefont {H.}~\bibnamefont
  {Ishino}},\ }\href {\doibase 10.1142/S2010194516601927} {\bibfield  {journal}
  {\bibinfo  {journal} {Int. J. Mod. Phys. Conf. Ser.}\ }\textbf {\bibinfo
  {volume} {43}},\ \bibinfo {pages} {1660192} (\bibinfo {year}
  {2016})}\BibitemShut {NoStop}%
\end{thebibliography}%

\end{document}